# Do We Expect More from Radiology AI than from Radiologists?

Maciej A. Mazurowski, PhD, MS, MA

From the Department of Radiology, Duke University, 2424 Erwin Rd, Suite 302, Durham, NC 27705. Received XXX; revision requested XXX; revision received XXX; accepted XXX; final version accepted XXX. **Address correspondence to** the author (e-mail: *maciej.mazurowski@duke.edu*).

The expectations of radiology AI do not match expectations of radiologists in terms of performance and explainability.

## Key Points

Expectations of radiology AI will guide its implementation in clinical settings.

The expectations of AI are based on a strong and justified mistrust about the way that AI makes decisions, but this mistrust is not mirrored in our expectations of human readers.

Expectations of radiologists differ from those of AI, particularly in terms of performance and explainability.

Conflicts of interest are listed at the end of this article.

What we expect from radiology AI algorithms will shape the selection and implementation of AI in the radiologic practice. In this paper I consider prevailing expectations of AI and compare them to expectations that we have of human readers. I observe that the expectations from AI and radiologists are fundamentally different. The expectations of AI are based on a strong and justified mistrust about the way that AI makes decisions. Because AI decisions are not well understood, it is difficult to know how the algorithms will behave in new, unexpected situations. However, this mistrust is not mirrored in our expectations of human readers. Despite well-proven idiosyncrasies and biases in human decision making, we take comfort from the assumption that others make decisions in a way as we do, and we trust our own decision making. Despite poor ability to explain decision making processes in humans, we accept explanations of decisions given by other humans. Because the goal of radiology is the most accurate radiologic interpretation, our expectations of radiologists and AI should be similar, and both should reflect a healthy mistrust of complicated and partially opaque decision processes undergoing in computer algorithms and human brains. This is generally not the case now.

## Current State of the Debate on the Role of AI in Radiology

With the advent of artificial intelligence (AI) algorithms that can perform image interpretation at the human level, recent years have brought a heated debate about what role AI will and should play in radiology. Within the community of radiologists, a prevailing sentiment is that AI will play a supportive role and it will not make a significant dent in the radiology workforce (1). However, there are significant voices within the broader radiology community that consider a disruption in the radiology workforce a possible or likely scenario (2,3).

I recently offered an analysis of the reasons why AI will not replace radiologists and I concluded that these arguments are far from conclusive (3). A reduction of image-interpretation jobs performed by radiologists is a real possibility. Why has it not happened then? Is it proof that the believers in AI replacing radiologists are wrong? Not at all. Or at least not yet. First, the science and practice of algorithm development and perhaps even more importantly, evaluation, need to take its course. Unfortunately, the community, encouraged by very strong initial results, formed unrealistic expectations of immediate clinical implementation. These expectations have not been met yet (4). One could compare this situation to seeing a therapeutic success of a drug in a mouse and expecting patients to be cured the following month. A proper process, particularly involving validation in real clinical scenarios, needs to take place. Unfortunately, rigorous statistical validation and clinical validation have not been traditionally a strength in the machine-learning community, but progress is being made.

There also may be a more important reason, beyond the time needed for development and validation, why AI algorithms are not yet a significant part of the clinical practice: Whether an algorithm will be implemented in a clinical practice depends on whether it serves the interests (often financial) of those who influence the decision of its implementation. This alignment of interests also affects the upstream development of the AI models. Those interests may be misaligned with the interests of patients. It is well-recognized that the incentives of many decision-makers in a health care system, such as insurers, hospitals, drug companies, or legislators, are not fully aligned or even may be highly misaligned with the interests of patients (5–7). These considerations also highly dependent on the available payment models which continue to change. Therefore, the insignificant extent to which AI is a part of radiology practices may not be a testament for the lack of algorithms that would benefit patients but rather a sign that those that intend to sell these algorithms have not yet found a way to plug themselves effectively into the complicated network of existing incentives and interests.

Careful consideration and sharing of opinions on how the future of AI in radiology should look and predictions on how it will look will affect lives of patients and therefore becomes an ethical matter. On the one hand, if the claims of AI replacing radiologists are exaggerated and they dissuade medical students from going into the radiology specialty (8), this may lead to shortage of physicians in this very important specialty (9). On the other hand, if AI really is to replace radiologists, medical students should indeed (at some point) start choosing other specialties to avoid the necessity of retraining and be able to use their talents in an optimal way. Furthermore, if AI can provide better health care to the patients, then an unfair opposition to its implementation among various stakeholders is unethical as it leads to worse patient outcomes (10). Therefore, a thoughtful consideration of these issues is crucial.

The opinions and decisions on whether AI algorithms should be a part of clinical practice are often justified, or at least rationalized, by stating that such algorithms do or do not meet certain expectations. In particular, one can hear criticism from AI skeptics listing issues such as AI being a black box, AI not having a "common sense," or simply not performing at a "radiologist level."

But an important question is rarely asked: Are our expectations of AI consistent with our expectations of radiologists? This article tackles that question.

## Explainable, Understandable Decisions

An important quality that we expect from radiology AI algorithms is transparency or explainability. There are some good reasons to expect it from decision makers (11). One reason is that we cannot test the algorithm across all possible conditions and against all possible pitfalls. Knowing the way decisions are made could increase the trust in the algorithm and build confidence that it will not make a mistake in unusual conditions. We also need to know that the algorithm does not unfairly discriminate based on sex, race, or other factors.

Explainability has taken different forms in machine learning models. Examples in deep learning include pointing to a part of the image that played a salient role in decision making (12) or pointing to the learned prototype images that look like the recognized object (13). In another approach, mutual information is used to add meaning or interpretability to the representation of the images in generative networks (14).

Even though attempts have been made to improve explainability, the most promising form of AI for image analysis-deep learning-is often perceived as a black box. And this perception is, to a large extent, correct because it is very challenging to answer the question of why a particular decision was made and not another.

Can a decision made by a radiologist be explained? The short answer is: not well. As in artificial neural networks, the decisions in a human brain are made through processing of input signals by a complicated network of interconnected neurons. We cannot perceive which neurons or systems of neurons fire at different times and, therefore, we have no mechanistic understanding of how individual decisions are made. Very little of this processing, at any level, rises to human consciousness. A complete mechanistic explanation would have to involve understanding the specific processes that took place when the individual decision was made, and such an explanation is not currently possible.

Can we provide an explanation of radiologists' decisions at a higher level using concepts such as shape, memory, or visual search? Psychology, cognitive science, and the overlapping discipline of image perception science offer various explanations of how humans make decisions and, in particular, how they interpret images. Although these fields offer some understanding of the cognitive processes involved in decision-making, they also inform us about how imperfect, biased, and riddled with inconsistencies our decision-making is (15–17). Radiology is not immune to these problems.

An example illustrates both the idiosyncratic nature of our decisions and our inability to get to the bottom of how we make individual decisions. In one study (18), the participants were asked the height of the Brandenburg Gate. Earlier, however, some of subjects were asked whether it was higher than a threshold of 150 m, and some were asked whether it was higher than a threshold of 25 m. Those that were "anchored" with the higher threshold estimated the Brandenburg Gate was almost 3 times higher. The experiment was repeated for a number of other guesses, including the age at which Gandhi died and the year of Leonardo da Vinci's birth. The results confirmed that such anchoring significantly impacts our decisions, creating a notable cognitive bias which may reduce the accuracy of our decisions without our realizing it. The anchoring bias is only one of dozens of biases that riddle our decision-making (16).

Additionally, while we are happy to offer the explanation of our individual decision-making or describe our "thought process," it turns out that we are not that good at introspection even at a higher level, and we may be quite wrong when pointing out the reasons for our decisions (17). Geoff Hinton (19) offers a great anecdote demonstrating our ability to explain visual interpretations. He points to a digit "2" and asks what digit it is. Once he hears the correct answer, he asks why. How was this decision made? Once we honestly ask ourselves this question, we have to admit that we simply do not know (although "If pushed, you will make up a story about how you did it," Hinton says).

Explanations of visual interpretations offered by radiologists resort to the assumption that we share common concepts such as shape or brightness, or more complicated concepts used in radiology such as nodule echogenicity in ultrasonography or mass margin in mammography. While this can be helpful, it may not reflect how a decision was actually made. Furthermore, significant experimental data on interreader variability (20,21) demonstrates that the assumption of these concepts being shared across radiologists is only partially true at best.

To summarize, our general scientific understanding and ability to explain decision making is limited at the conceptual level and even poorer when it comes to the mechanistic level. To the extent that we do understand human decision-making, we know that it is a mixture between what we would consider perceptive and reasonable and what we would consider idiosyncratic and biased. And finally, even if we have some understanding of decision-making in humans in general, our ability for introspection and accurate explanation of individual decisions that we make is moderate at best.

Despite these significant limitations, explanations of radiology decisions offered by radiologists who make them can still have two important functions: (1) they can help educate radiologists in training; and (2) they can improve confidence in individual decisions by allowing others to come to the same conclusion. However, such explanations cannot guarantee that a radiologist will not make an error in unseen and unusual scenarios. Despite the lack of such guarantee for radiologists, explanations of such quality are often expected from AI. It appears that radiologists and AI systems, for slightly different reasons, are in the same boat. Neither can offer a high level of reassurance through explanations of their decisions.

## Common Sense and Intuition

A colleague giving a keynote lecture brought up the Hudson River landing of the US Airways Flight 1549 with 150 passengers and 3 flight attendants on board. The speaker concluded that the passengers were lucky that it was a human flying the plane because an algorithm would lack the ability to make a "judgment call" needed to make the right decision. The speaker asked whether we would rather safely land with a human pilot or die in a plane controlled by AI. You can imagine the answer to this ill-posed question.

This reasoning is incorrect in a number of ways, and this particular example really backfires. First, given our current understanding of this specific landing, the pilot had virtually no chance to land successfully at an airport exactly because he, a human, needed time (35 seconds) to assess the situation. A good algorithm could potentially make this decision in less than 1 second and, according to simulations, would have had about a 50% chance to land at a nearby airport. Second, the successful outcome does not mean that a good decision was made. Rather, an overall probability of success should be considered. Finally, and most importantly for this discussion, the speaker implied that to make the correct decision, one needs to be able to make a "judgment call."

The inability of AI to exercise common sense, make judgment calls, or have intuition appears to be a common worry among AI skeptics. The reason for the worry is that based on our

Page 4 of 8

introspection, these concepts play an important role in decision-making. They make sense to us subjectively and also appear to be understood intersubjectively. We believe that others must undergo similar processes internally.

While on the surface very important, their significance fades away when investigated closely. Judgment calls appear to be decisions that are made with insufficient data on the potential outcomes of our decisions. Intuition appears to be a justification for decisions/opinions where we struggle to access good arguments for it. Common sense is an intuition that is hoped to be shared by many. Decisions can be made by algorithmic processing of data and in the presence of poor information and some of such processing could be called "judgment calls," or "common sense." There is no need for "judgment calls," "common sense," or "intuition" in a sense that goes beyond such processing.

## AI Needs to Be as Good or Better than Radiologists

Algorithm performance is perhaps the least controversial of the expectations of the AI algorithms. It is quite clear that an AI algorithm should not make too many mistakes if it were to be used in a clinical setting with different performance being acceptable for different clinical tasks. We expect that AI will operate with radiologist-level (or better) performance. But what does that mean? Which radiologist? What metric should we use? How is the ground truth established? And, importantly, are all radiologists operating at "radiologist level"?

A challenging question in this context is: Why do we require AI to match the performance of certain radiologists and we do not require all radiologists to match the same performance?

A typical way to assess an algorithm is to have the algorithm and a group of radiologists to perform the same task and assess performance of the AI and the radiologists against some prespecified gold standard. The expectation from AI is that the algorithm's performance will match or exceed the *average or median* performance of the radiologists. As long as the radiologists participating in the experiments are representative of the population of radiologists for a given task, this is a fair way to compare AI to radiologists as a group. However, it is not fair when comparing AI to individual radiologists. The expectation from AI is that it matches the average radiologists' performance. There is no such expectation from individual radiologists. In fact, if there was, about half of them would not meet such expectation and while there are outliers showing almost superhuman excellence in image interpretation, there are also outliers showing performance that is worrisome. Could the radiologists that perform better than average continue performing the specific task and those below average be reassigned to a different task and have AI take over their duty (assuming that AI matches performance of an average radiologist)? Mathematically, this would certainly lead to better overall performance.

There is a fundamental disparity between the expectations for performance, established through thorough testing, for AI and for radiologists. Radiologists are rarely tested in clinical scenarios, with large numbers of cases, and an established gold standard. Moreover, some radiologists perform below the level expected from AI (ie, the level of an average radiologist) and this is widely accepted as interreader variability status quo. This is not to say that we should not test AI since we rarely test radiologists. We should test AI and we should do it rigorously. However, we should take a step back and stress that the question that needs to be answered is what confidence do we have that introducing AI to a radiologic practice will improve the lives of the patients? And this question should be determined based on the rigorous assessment of both AI and radiologists and not on the trust that we place in our fellow human beings.

## Conclusions

There is a common thread in the discussions above. We expect more from AI than from humans because we understand and trust AI less. Why do we extend trust toward other human beings that we do not extend toward algorithms? The likely answer is because we feel that we know how we, as humans, perceive and make decisions, and we find comfort in thinking that as long as others make decisions in a similar way and undergo similar training, they will do just fine. Or at least they will avoid catastrophic mistakes ascribed to AI such as mistaking fried chicken for puppies. But this may be a false comfort. Our limited understanding of human decision-making process is not enough to ensure avoidance of catastrophic mistakes.

On the one hand, we should, to a reasonable extent, mistrust AI to make correct decisions regardless of the setting. Therefore, it is crucial that we test the algorithms in real-world scenarios and on datasets that are diverse in terms of patient populations, scanner parameters, imaging protocols, technologists who acquire the images, and any other parameters that may reasonably affect the decision.

On the other hand, putting an undue burden on the algorithms because we trust ourselves as human beings may not be a correct policy. The goal should be to implement a radiologic practice that provides the most benefit to the patients. This needs to be based on a reasonable mistrust toward the algorithm, but it should not be based on an unjustified trust toward our own human ability to perceive and make decisions.

**Acknowledgments:** The author would like to thank Edward Cohen and Walter Wiggins, MD, for their valuable discussions and edits to this article.

**Disclosures of Conflicts of Interest: M.A.M.** Activities related to the present article: disclosed no relevant relationships. Activities not related to the present article: disclosed no relevant relationships. Other relationships: institution has patent pending (US patent app. 15/209,212) for systems and methods for extracting prognostic image features.